\documentclass[prc,aps,preprint,showpacs]{revtex4}
\usepackage{graphicx}
\newcommand{\simgeq}{\; \raisebox{-0.4ex}{\tiny$\stackrel
{{\textstyle>}}{\sim}$}\;}
\newcommand{\simleq}{\; \raisebox{-0.4ex}{\tiny$\stackrel
{{\textstyle<}}{\sim}$}\;}

\begin{document}

\title{Shell structure of weakly-bound and resonant neutrons}

\vspace{1cm}

\author{ Ikuko Hamamoto$^{1,2}$ }

\address{
$^{1}$ {\it Division of Mathematical Physics, Lund Institute of Technology 
at the University of Lund, Lund, Sweden}   \\
$^{2}$ {\it The Niels Bohr Institute, Blegdamsvej 17, 
Copenhagen \O,
DK-2100, Denmark} \\ 
}


\vspace{2cm}

\begin{abstract}
The systematic 
change of shell structure in both weakly bound and resonant neutron
one-particle levels in nuclei 
towards the neutron drip line is exhibited,
solving the coupled equations derived 
from the Schr\"{o}dinger equation in coordinate space with the correct 
asymptotic behaviour of wave functions for $r \rightarrow \infty$.
The change comes from the behaviour unique in the one-particle motion with low
orbital angular momenta compared with 
that with high orbital angular momenta. 
The observed 
deformation of very neutron-rich nuclei with $N \simgeq 20$ in the island of
inversion is a natural
result of this changed shell structure, while a possible deformation of
neutron-drip-line nuclei with $N \approx 51$, which are not yet observed,  
is suggested.

\end{abstract}


\maketitle

\section{Introduction}
The study of one-particle motion in deformed potentials is the basis for
understanding the structure of deformed drip-line nuclei.  Since the Fermi level
of drip line nuclei lies close to the continuum, both weakly-bound and
positive-energy one-particle levels play a crucial role in the many-body
correlation of those nuclei.  Among an infinite number of one-particle levels at
a given positive-energy, only some selected levels related to one-particle
resonant levels will be important to bound states of drip-line nuclei.  
The motion of weakly-bound or low-energy resonant protons is not so
different from that of well-bound protons except for very light nuclei, 
due to the presence of the Coulomb
barrier.  Therefore, the present talk is devoted to the change of 
shell structure of one-neutron levels. 

The change of the shell structure 
in weakly-bound 
and resonant one-neutron levels is exhibited, which comes from 
the behavior unique in the one-particle motion with 
low orbital-angular-momenta ($\ell$). 
For spherical shape the centrifugal barrier is absent 
for $\ell$=0 orbits while the barrier is low for $\ell$=1.  
For deformed shape a given one-particle wave-function contains 
components with various $\ell$ values, 
however, the minimum value of possible $\ell$ for the
wave-function components plays a crucial role in the present problem. 
For example, in the case of 
axially-symmetric quadrupole deformation $\ell_{min}$= 0 
for $\Omega^{\pi}$ = $1/2^+$ and $\ell_{min}$=1 for both 
$\Omega^{\pi}$ = 3/2$^-$ and 1/2$^-$, where $\Omega$ expresses the component of
the particle angular-momentum along the symmetry axis.  
For bound one-particle levels the $\ell = 0$ component becomes always 
overwhelming in one-particle 
wave-functions with $\Omega^{\pi} = 1/2^+$, as
the binding energy approaches zero. 
Namely, in the limit of small binding energy  
the halo phenomena which are created exclusively 
by the $\ell = 0$ neutron wave-function appear 
for both spherical and deformed shapes. 
On the other hand, the decay of
$\Omega^{\pi} = 1/2^+$ one-particle resonances goes most easily through the 
$s_{1/2}$ channel because of the absence of the centrifugal barrier.
Therefore, it is in general difficult for an $\Omega^{\pi}$ = 1/2$^+$ Nilsson
level to survive as a resonant level in the positive-energy region. 
Nevertheless, we note that 
a given one-particle resonance with $\Omega^{\pi} = 1/2^+$ may survive even at
several MeV excitation-energy, if the higher-$\ell$ components 
are predominant in the
wave function inside the potential.  In such cases the existence of
the resonance is made possible by the presence of 
the higher-$\ell$ components while
the width of the resonance is determined by the small $\ell = 0$ component.  

One-particle resonant levels for spherical shape 
are defined in terms of phase shift as described in standard textbooks. 
On the other hand, the definition of one-particle resonance 
for deformed potentials is obtained using the 
eigenphase formalism \cite{RGN66,IH05,IH06}. 
The definition is a natural extension of the definition of one-particle
resonance for spherical potentials in terms of phase shift.

Taking the three bound one-particle levels for neutrons in the $sd$ shell, 
in \cite{IH10-g} it is shown how the order of the three levels varies as a
function of binding energies: from the bottom to the top in energy, 
(i) the relative level order is 
$1d_{5/2}$, $1d_{3/2}$ and $2s_{1/2}$ for strongly bound cases,; 
(ii) $1d_{5/2}$, $2s_{1/2}$ and $1d_{3/2}$ for about 10 MeV binding and 
(iii) $2s_{1/2}$, $1d_{5/2}$ and $1d_{3/2}$ for very weak bindings. 
Namely, 
the $\ell = 0$ level shifts downwards relative to the $\ell = 2$
levels, as the potential strength becomes weaker or the binding energy becomes 
smaller.  
The level order 
in (ii) is known as the usual energy splitting in stable $sd$-shell nuclei, 
which is used in the shell model. 
In the present talk we concentrate on the shell structure of 
bound and resonant neutron levels in the $pf$-shell
as well as in the $50 < N \leq 82$ major shell, taking examples of
a stable $sd$-shell nucleus (figure 1), a known deformed neutron-drip-line 
nucleus (figure 2), 
and a not yet
observed neutron-drip-line nucleus in the region of medium mass (figure 3).

\section{Model and numerical examples}
In the description of one-particle resonances the formalism using 
complex energy is sometimes popular 
in the literature.  However, it is equally possible to formulate the resonances
using real energy.  I prefer to choosing the latter, since the treatment of
certain observable quantities such as matrix-elements or 
wave-function components is much simpler
with real variables. 

Solving the coupled equations derived from 
the Schr\"{o}dinger equation in coordinate space with the correct 
asymptotic behaviour of wave functions for $r \rightarrow \infty$, 
the shell structure in axially symmetric
quadrupole-deformed Woods-Saxon potentials is studied. Whereas one-particle
bound levels in the deformed potential are obtained by solving the well-known
eigenvalue problem, for positive-energy one-particle levels all eigenphases 
for a given $\Omega^{\pi}$ are calculated, 
and the possible one-particle resonant level 
for a given energy $\varepsilon_{\Omega}$ and a given $\Omega^{\pi}$ value may
be  
identified, which is associated with a particular eigenphase among all possible 
eigenphases.  One-particle resonance is not obtained if none of calculated
eigenphases do not increase through $\pi / 2$ as the one-particle energy
increases at the given energy.  
The relevant formulation can be found in \cite{IH05,IH06}. 

The parameters of the Woods-Saxon potential are chosen as follows: the
diffuseness $a$= 0.67 fm and the radius $R = r_0 A^{1/3}$ with $r_0$=1.27 fm, 
while the depth which depends on the neutron excess in respective nuclei 
is taken approximately from
the expression (2-182) in \cite{BM69}.  The neutron potential 
for nuclei with a neutron excess is shallower than that for $N=Z$ nuclei.

\subsection{Nilsson diagram for a stable nucleus $^{25}_{12}$Mg$_{13}$}
The analysis of observed spectroscopic properties of low-lying states in light
deformed mirror nuclei, $^{25}_{12}$Mg$_{13}$ and $^{25}_{13}$Al$_{12}$, in
terms of one-particle motion in a deformed potential is very successful 
\cite{BM75} using the deformation parameter $\delta_{osc} \approx 0.4$.
In \cite{BM75} the Nilsson diagram based on a modified oscillator
potential is used.  
Since the neutron separation energy $S_n$ of $^{25}$Mg is 7.3 MeV,
the properties of low-lying levels could well be described using modified
oscillator potential when the related parameters were properly adjusted. 
In figure 1 the Nilsson diagram based on a deformed Woods-Saxon potential is
shown.  
Observed spectroscopic properties of low-lying states are equally well 
understood also using the Nilsson diagram in figure 1. 
  
At $\beta$=0 of figure 1 
the $2s_{1/2}$ level appears around the middle of the $1d_{5/2}$
and $1d_{3/2}$ level, as usually assumed.  
A striking difference of figure 1 from the Nilsson diagram based on the
modified oscillator potential (or from one-particle energies 
used in the conventional shell model) is seen 
in the energies of the $2p_{3/2}$ and 
$1f_{7/2}$ levels.
Calculated one-particle resonant energies are 
$\varepsilon _{res}(p_{3/2})$ = +0.31 MeV 
and $\varepsilon_{res}(f_{7/2})$ = +0.32 MeV.  Namely, the $p_{3/2}$ level lies
even lower than the $f_{7/2}$ level. In other words, the $N$=28 energy gap
totally disappears in the low-energy resonance spectra.

\subsection{Nilsson diagram for a deformed neutron-drip-line nucleus 
$^{31}_{10}$Ne$_{21}$}
In figure 2 the Nilsson diagram appropriate for neutrons in 
$^{31}_{10}$Ne$_{21}$, of which
the measured neutron separation energy 
$S_n$ is 0.29$\pm$1.64 MeV, is shown.
At $\beta$=0 the $f_{7/2}$ one-neutron resonance is found at 2.04 MeV, while
neither $p_{3/2}$ nor $p_{1/2}$ resonance defined by the eigenphase formalism 
is obtained. However, the $p_{3/2}$ resonance lying clearly lower than 
the $f_{7/2}$ resonance is found if we use a slightly more attractive
Woods-Saxon potential.  The next low-lying one-particle resonant level 
for $\beta$=0 is the $f_{5/2}$ level at 9.50 MeV. 
  
The measured very low excitation-energies of the first 2$^+$ state of both 
$^{30}$Ne and $^{32}$Ne indicate that the Ne isotope with the neutron numbers 
is deformed.  Furthermore, the large Coulomb breakup cross section reported in
\cite{TN09} clearly suggests the halo nature of the ground state of
$^{31}$Ne.  
In the conventional shell model the 21st neutron for a spherical shape  
occupies the $1f_{7/2}$ shell and, thus,  
the related halo phenomena are impossible to obtain.  
If $^{31}$Ne is prolately deformed, from figure 2 we obtain 
the bound Nilsson level which is to be occupied 
by the 21st neutron having a halo structure: [330 1/2] for $0.20 \simleq \beta
\simleq 0.30$ and [321 3/2]for $0.40 \simleq \beta \simleq 0.58$.  In both cases 
the ground state should have 3/2$^-$ and the $p$-wave neutron halo together 
with the deformed core $^{30}$Ne. The measurement of $S_n$ with an accuracy 
which is an order of magnitude better than the presently available one 
will clarify the intrinsic configuration of the ground state of $^{31}$Ne 
\cite{IH10a}. 

Examining figure 2 we may also note that the presence of $^{31}$Ne inside the
neutron drip line is possibly realized by the deformation, which is created by
the Jahn-Teller effect due to the near degeneracy of $1f_{7/2}$, $2p_{3/2}$ and 
$2p_{1/2}$ shells in the continuum for the spherical shape.  
Furthermore, it is remarked that 
the measured magnetic moment of the isotone nucleus 
$^{33}_{12}$Mg$_{21}$ \cite{DY07} is in good agreement with the estimated value
for the prolately deformed shape \cite{DY10}.

\subsection{Nilsson diagram for neutron-drip-line nuclei with $A \approx 75$ and
$N \approx 50$}
In figure 3 the Nilsson diagram appropriate for neutron-drip-line nuclei 
with $N \approx 51$ (for example, $^{75}_{24}$Cr$_{51}$) is shown. 
Since in this example the $1f_{7/2}$, $2p_{3/2}$, $1f_{5/2}$ and $2p_{1/2}$ 
levels are well bound, the presence of the magic number $N = 28$ which is
established for stable nuclei is clearly seen.  Moreover, 
the level order of four levels in the $1f-2p$ major shell is exactly what is
used in traditional shell-model calculations. 
In contrast, we note the peculiar order 
of weakly-bound and resonant one-particle 
levels which belong to the $50 < N \leq 82$ major shell. 
At $\beta = 0$ the very weakly bound $3s_{1/2}$ level is almost degenerate with 
(but slightly lower 
than) the $2d_{5/2}$ level, and the one-particle resonant levels in 
the $50 < N \leq 82$ major shell which are not plotted in the figure are:
$\varepsilon_{res}(1g_{7/2}) = +3.44 MeV$ and 
$\varepsilon_{res}(1h_{11/2}) = +5.48 MeV$.  
It is noted that the level order in the major shell consisting of 
the $3s_{1/2}$, $2d_{3/2}$, $2d_{5/2}$ $1g_{7/2}$, and 
$1h_{11/2}$ levels is totally different from what is known from stable nuclei.
  
A similarity of the level bunching 
around $N$=21 in figure 2 to that around  $N$=51 in figure 3 
as a function of deformation is seen 
especially for $\beta \geq 0$.  Considering the fact 
that the $N$=21 neutron-drip-line nuclei, $^{31}$Ne and $^{33}$Mg, 
are deformed, 
the nucleus such as $^{75}_{24}$Cr$_{51}$ may be also deformed, 
if it indeed lies
inside the neutron drip line.  
In other words, there might be $N=51$ nuclei such that 
they may remain inside the neutron drip line due to the possible deformation. 
Furthermore, instead of having $I^{\pi}$ = 7/2$^+$ or 5/2$^+$ coming from
the occupancy of 
$1g_{7/2}$ or $2d_{5/2}$ neutron orbits, respectively,  
by the 51st neutron 
the neutron-drip-line nuclei with $N=51$ have a good chance to have the
ground or very low-lying $I^{\pi}$ = 1/2$^+$ state, irrespective of 
spherical or deformed shape.

\section{Summary and discussions}
It is known that the presence of the surface in realistic potentials of stable
nuclei pushes down one-particle levels with higher $\ell$ relative to lower
$\ell$ levels among the levels belonging to a given major shell of the 
harmonic-oscillator potential.  
On the other hand, weakly-bound neutrons with small $\ell$ have an appreciable
probability of being outside the potential compared with those with larger 
$\ell$ and thus are insensitive to the strength of the potential.  Consequently,
those small-$\ell$ levels are pushed down relative to high-$\ell$ levels, as the
binding energy becomes small.  The resulting shell structure, which is
quite different from what we are accustomed to in stable nuclei, would appear in
weakly-bound and resonant neutron one-particle levels and affect the nuclear
many-body problem in neutron drip line nuclei.  
In the present talk I have chosen to demonstrate the shell structure 
of one-particle 
levels in the $1f$-$2p$ shell as well as in the $50 < N \leq 82$ major 
shell, when the levels become weakly-bound or resonant.   
Examining the obtained shell structure one may understand 
the possible origin of the deformation of nuclei in 
the island of inversion ($N \simgeq 20$) 
as well as a possible deformation of neutron drip line
nuclei with $N \approx 51$.

The systematic change of shell structure described in the present article 
is strictly related to the characteristic behavior of 
the one-particle motion with small orbital 
angular-momenta.  The change cannot be estimated by using harmonic-oscillator
potential (or wave function).
In the present meeting the change of shell structure of neutrons (protons) as
the proton (neutron) number varies is discussed, considering the tensor force 
between protons and neutrons using harmonic-oscillator wave functions. 
Though some of the shell-structure change looks numerically similar to that
shown in the present work, the relevant physics is essentially
different.


\mbox{}

\pagebreak[4]

\newpage

\begin{figure}
\begin{center}
\includegraphics[width=12cm]{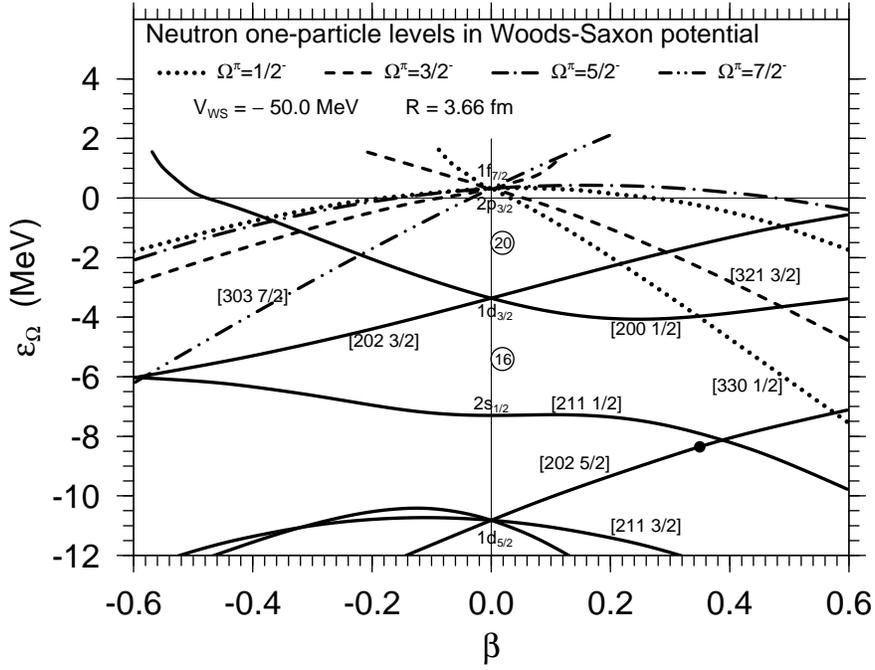}
\vspace{10cm}
\caption{Neutron one-particle levels as a
function of axially-symmetric quadrupole deformation.  The depth and radius of
the Woods-Saxon potential are designed approximately 
for $^{25}_{12}$Mg$_{13}$. The possible occupancy of the 13th neutron in the
ground state of $^{25}$Mg is indicated by the filled circle.  Some one-particle
levels are denoted by the asymptotic quantum numbers, [$N n_z \Lambda \Omega$].
One-particle levels plotted for $\varepsilon_{\Omega} > 0$ are resonant levels 
obtained by using the eigenphase formalism, of which the widths are not plotted
for simplicity.  Note that one-particle resonant $p_{3/2}$ and
$f_{7/2}$ levels, of which the resonant energies are about 0.3 MeV, 
are almost degenerate. }
\label{fig1}
\end{center}
\end{figure}

\newpage
\begin{figure}
\begin{center}
\includegraphics[width=12cm]{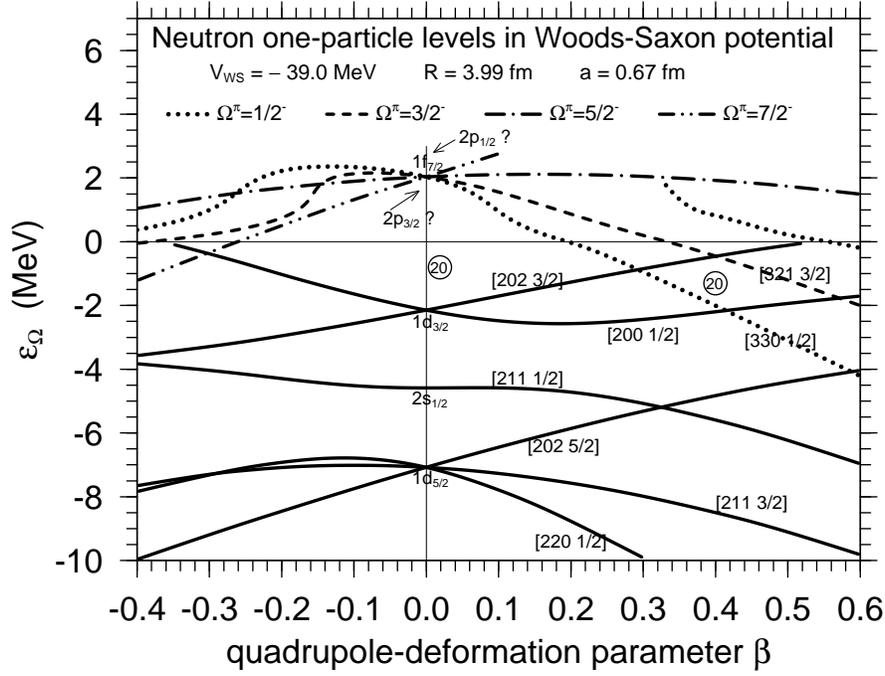}
\vspace{10cm}
\caption{Neutron one-particle levels as a
function of axially-symmetric quadrupole deformation.  The depth and radius of
the Woods-Saxon potential are designed approximately 
for $^{31}_{10}$Ne$_{21}$. The neutron number 20, which is obtained by filling
in all lower-lying levels, is indicated with a circle.  The approximate
positions of the $2p_{3/2}$ and $2p_{1/2}$ levels for $\beta = 0$ are indicated
with ''?'', which are extrapolated from the resonant energies obtained 
by using a slightly more attractive Woods-Saxon potential. For the potential 
with the present strength the $p_{3/2}$ and $p_{1/2}$ resonant levels are not 
obtained for $\varepsilon_{\Omega} >$ 1.3 MeV, due to the low 
centrifugal barrier for $\ell$=1 orbits. The calculated widths of one-particle
resonant levels are not plotted, for simplicity.  }
\label{fig2}
\end{center}
\end{figure}

\newpage
\begin{figure}
\begin{center}
\includegraphics[width=12cm]{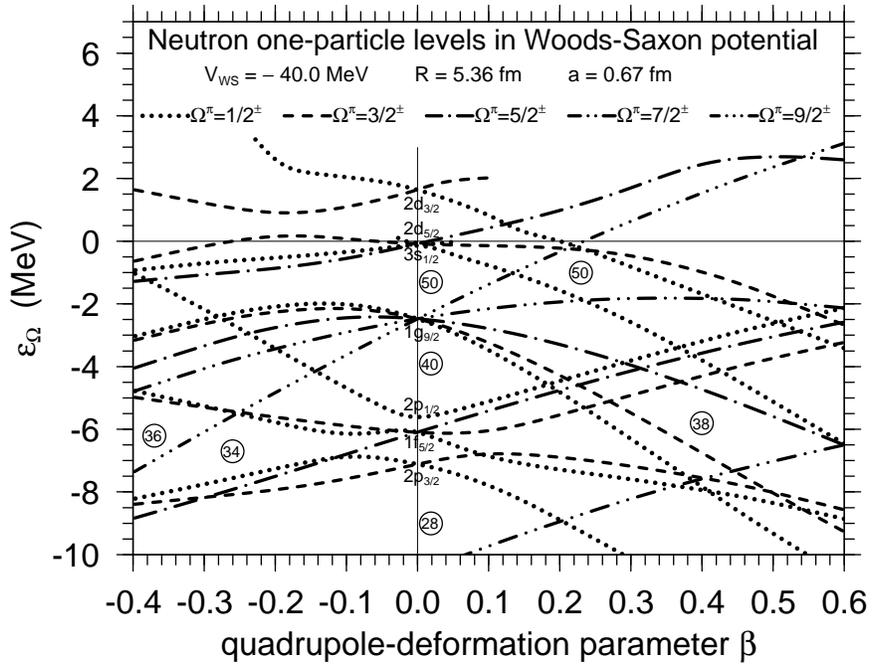}
\vspace{10cm}
\caption{Neutron one-particle levels as a
function of axially-symmetric quadrupole deformation.  The depth and radius of
the Woods-Saxon potential are designed approximately 
for $^{75}_{24}$Cr$_{51}$.}
\label{fig3}
\end{center}
\end{figure}

\end{document}